\documentclass[a4paper,12pt, epsfig]{article}
\def\letter{0}\def\pr{0}
\usepackage{epsfig}
\usepackage{epstopdf}
\usepackage{graphicx}
\usepackage{ifthen}

\usepackage{amsmath}
\usepackage[psamsfonts]{amssymb}
\usepackage{euscript}

\usepackage{latexsym}
\usepackage[arrow,matrix,curve]{xy}

\newskip\humongous \humongous=0pt plus 1000pt minus 1000pt

\newif\ifdtup

\def\,{\hspace{-.1cm}}
\def\hsp{,\hspace{.7cm}}

\def\fc#1#2 {\frac{n}{q}#1\frac{n}{q}#2}

\def\kt{\kappa}
\def\kt{\mathfrak{K}}
\def\ks{|\kt\rangle}
\def\bdk{B^\dag_\kt}

\newcommand{\vac}{\ensuremath{|0\rangle}}

\renewcommand{\theequation}{\arabic{section}.\arabic{equation}}
\renewcommand{\(}{\begin{equation}}
\renewcommand{\)}{end{equation} \vspace{-.05in}\linebreak}

\newcounter{saveeqn}
\newcounter{savealpheqn}

\newcommand{\alpheqn}{\setcounter{saveeqn}{\value{equation}}%
  \stepcounter{saveeqn}\setcounter{equation}{0}%
  \renewcommand{\theequation}{\mbox{\arabic{section}.\arabic{saveeqn}
\alph{equation}}}
  \renewcommand{\)}{\end{equation}}}
\def\part#1{\frac{\partial}{\partial{#1}}}%
\def\group#1{\refstepcounter{equation}\setcounter{saveeqn}
 {\value{equation}}%
  \label{#1}\setcounter{equation}{0}%
\renewcommand{\theequation}{\mbox{\arabic{section}.\arabic{saveeqn}
\alph{equation}}}
  \renewcommand{\)}{\end{equation}}}
\newcommand{\reseteqn}{\setcounter{equation}{\value{saveeqn}}%
  \renewcommand{\theequation}{\arabic{section}.\arabic{equation}}%
  \renewcommand{\)}{\end{equation}}}

\newcommand{\aalpheqn}{\setcounter{saveeqn}{\value{equation}}%
  \stepcounter{saveeqn}\setcounter{equation}{0}%
  \renewcommand{\theequation}{\mbox{
        \Alph{subsection}.\arabic{saveeqn}\alph{equation}}}
   \renewcommand{\)}{\end{equation}}}
\newcommand{\areseteqn}{\setcounter{equation}{\value{saveeqn}}%
  \renewcommand{\theequation}{\Alph{subsection}.\arabic{equation}}%
  \renewcommand{\)}{\end{equation}}}

\renewcommand{\thefootnote}{\alph{footnote}}
\renewcommand{\(}{\begin{equation}}
\renewcommand{\)}{\end{equation}}
\newcommand{\ba}{\begin{eqnarray}}
\newcommand{\ea}{\end{eqnarray}}
\newcommand{\cbp}{\mathop{\vtop{\ialign{##\crcr
   $\hfil\displaystyle{}\hfil$\crcr\noalign{\kern-13pt\nointerlineskip}
   \BIG{)}\hskip0pt\crcr\noalign{\kern3pt}}}}}
\newcommand{\pa}{\mathop{\vtop{\ialign{##\crcr

$\hfil\displaystyle{\oplus}\hfil$\crcr\noalign{\kern+1pt\nointerlineskip
}
   \hspace{.08in}$^{\alpha=0}$\hskip6pt\crcr\noalign{\kern3pt}}}}}
\renewcommand{\hsp}{,\hspace{.3in}}
\newcommand{\p}{^\prime}
\newcommand{\pp}{^{\prime\prime}}

\catcode`\@=11
\def\vereq#1#2{\lower3pt\vbox{\baselineskip1.5pt \lineskip1.5pt
\ialign{$\m@th#1\hfill##\hfil$\crcr#2\crcr\sim\crcr}}}
\catcode`\@=12

\renewcommand{\(}{\begin{equation}}
\renewcommand{\)}{\end{equation}}

\def\pin#1{\int \frac{d#1}{2\pi}}
\def\pink#1{\int \frac{d^{#1}k}{(2\pi)^{#1}}}

\def\pinkp#1{\int \frac{d^{#1}k\p}{(2\pi)^{#1}}}

\def\Bd#1{B^\dag_{k_{#1}}}

\def\df{\mathcal{D}_f}
\def\B#1{B^\dag_{k_{#1}}}

\def\I{\mathcal{I}}
\def\vl#1#2#3{\vector(#1,#2){#3}\line(#1,#2){#3}}
\def\wk#1{\omega_{k_{#1}}}
\def\wkp#1{\omega_{k^{'}_{#1}}}

\newcommand{\beas}{\begin{eqnarray*}}
\newcommand{\eeas}{\end{eqnarray*}}

\newcommand{\bquo}{\begin{quote}}
\newcommand{\enqu}{\end{quote}}

\def\ch{{\mathcal{H}}}

\def\ok#1{\omega_{k_{#1}}}
\def\okp#1{\omega_{k\p_{#1}}}
\def\V#1{V^{(#1)}[gf(x)]}

\newcommand{\beq}{\begin{equation}}
\newcommand{\eeq}{\end{equation}}
\newcommand{\bea}{\begin{eqnarray}}
\newcommand{\eea}{\end{eqnarray}}

\newskip\humongous \humongous=0pt plus 1000pt minus 1000pt

\newif\ifdtup

\catcode`\@=11

\ifthenelse{\equal{\letter}{0}}{ 

\@addtoreset{equation}{section}
\def\theequation{\arabic{section}.\arabic{equation}}

\def\@normalsize{\@setsize\normalsize{15pt}\xiipt\@xiipt
\abovedisplayskip 14pt plus3pt minus3pt%
\belowdisplayskip \abovedisplayskip
\abovedisplayshortskip \z@ plus3pt%
\belowdisplayshortskip 7pt plus3.5pt minus0pt}

\def\small{\@setsize\small{13.6pt}\xipt\@xipt
\abovedisplayskip 13pt plus3pt minus3pt%
\belowdisplayskip \abovedisplayskip
\abovedisplayshortskip \z@ plus3pt%
\belowdisplayshortskip 7pt plus3.5pt minus0pt
\def\@listi{\parsep 4.5pt plus 2pt minus 1pt
      \itemsep \parsep
      \topsep 9pt plus 3pt minus 3pt}}

\relax

\def\section{\@startsection{section}{1}{\z@}{3.5ex plus 1ex minus  .2ex}{2.3ex plus .2ex}{\large\bf}}

\def\thesection{\arabic{section}}
\def\thesubsection{\arabic{section}.\arabic{subsection}}

\def\appendix{\setcounter{section}{0}
 \def\thesection{Appendix \Alph{section}}
 \def\thesubsection{\Alph{section}.\arabic{subsection}}
 \def\theequation{\Alph{section}.\arabic{equation}}}
\renewcommand{\theequation}{\arabic{section}.\arabic{equation}}

}{
\renewcommand{\theequation}{\arabic{equation}}

} 

\begin{document}
\def\thefootnote{\fnsymbol{footnote}}
\def\thetitle{Excited Kinks as Quantum States}
\def\autone{Jarah Evslin}
\def\auttwo{Hengyuan Guo}
\def\affa{Institute of Modern Physics, NanChangLu 509, Lanzhou 730000, China}
\def\affb{University of the Chinese Academy of Sciences, YuQuanLu 19A, Beijing 100049, China}

\title{Titolo}

\ifthenelse{\equal{\pr}{1}}{
\title{\thetitle}
\author{\autone}
\author{\auttwo}
\affiliation {\affa}
\affiliation {\affb}

}{

\begin{center}
{\large {\bf \thetitle}}

\bigskip

\bigskip


{\large \noindent  \autone{${}^{1,2}$} and \auttwo{${}^{1,2}$}}


\vskip.7cm

1) \affa\\
2) \affb\\

\end{center}

}

\begin{abstract}
\noindent
At one loop, quantum kinks are described by a sum of quantum harmonic oscillator Hamiltonians, and so their spectra are known exactly.  We find the first correction beyond one loop to the quantum states corresponding to kinks with an excited bound or unbound normal mode, and also the corresponding two-loop correction to the energy cost of exciting the normal mode.  In the case of unbound normal modes, this correction is equal to sum of the corresponding nonrelativistic kinetic energy plus the usual one-loop correction to the mass of the corresponding plane wave in the absence of a kink.  We also sketch a diagrammatic method for such calculations.


\noindent

\end{abstract}

%
\setcounter{footnote}{0}
\renewcommand{\thefootnote}{\arabic{footnote}}

\ifthenelse{\equal{\pr}{1}}{
\maketitle
}{}

\section{Introduction}

The scattering of kinks is a major industry.  It has a long history, with quantum kink scattering already in Refs.~\cite{christlee75,goldstone}.  However quantum kink scattering has proved to be cumbersome and so the most interesting phenomenology \cite{mak78,moshir81,adamwall,lankink} has only been revealed classically.  Classically a key role in the resonance phenomenon \cite{campres}, spectral walls \cite{adamwall} and even wobbling kink multiple scattering \cite{alonso2020} appears to be played by bound normal modes.  However the exact role played by these modes is unclear, as the resonances have been observed in kinks with no bound normal modes \cite{takyi}.   These modes themselves enjoy a rich phenomenology.  They can  be excited by external perturbations \cite{quintero2000} and they can store energy from a collision~\cite{campres}.

Clearly it would be of interest to understand these phenomena in the full quantum theory.   At one loop the exact spectrum of quantum kinks is known \cite{dhn2}, as kinks are simply described by quantum harmonic oscillators for each normal mode together with a free quantum particle describing the center of mass. 

Recently \cite{me2looplett} a method was proposed which allows the practical calculation of higher-loop states.  This method, to be reviewed in Sec.~\ref{revsez}, constructs a kink sector Hamiltonian $H\p$ and momentum $P\p$ via a unitarity transformation of the defining Hamiltonian $H$ and momentum $P$.  Then states can be pushed beyond one loop by first imposing perturbatively that they be eigenstates of the momentum $P\p$, which fixes the state up to a few coefficients, and then applying old-fashioned perturbation theory in $H\p$ to fix these remaining coefficients.  The corresponding eigenstates of $H$ and $P$ are recovered from this result via the inverse unitary transformation.

So far this method has only been applied to the kink ground state.  However, in light of the above motivation, in the present paper we will apply it to kinks excited by a single continuum or bound normal mode, in their center of mass frame.  We will find the first correction to the states beyond one loop and also will find the corresponding two-loop mass correction.  With these states in hand, it will be possible in future work to compute their form factors and matrix elements, which in turn may be applied to compute fully quantum scattering amplitudes.   While the one-loop form factors have long been known to be simply related to the classical kink solutions \cite{goldstone}, it will be clear that at next order many matrix elements that vanish at one loop no longer vanish, presumably leading to novel physical effects in quantum scattering. 

In Sec.~\ref{statsez} we will construct the leading order correction to the one-loop states corresponding to quantum kinks with excited continuum or discrete normal modes.   In Sec.~\ref{masssez} we will find the corresponding two-loop mass shifts.  Finally in Sec.~\ref{dessinsez} we will sketch a diagrammatic method to perform such calculations in general.  The main notation is summarized in Table~\ref{notab}.  In \ref{app} we check that our state satisfies the most constraining component of the Schrodinger equation, which summarizes the condition that it be a Hamiltonian eigenstate.

\section{Review} \label{revsez}

\begin{table}
\begin{tabular}{|l|l|}
\hline
Operator&Description\\
\hline
$\phi(x),\ \pi(x)$&The real scalar field and its conjugate momentum\\
$A^\dag_p,\ A_p$&Creation and annihilation operators in plane wave basis\\
$B^\dag_k,\ B_k$&Creation and annihilation operators in normal mode basis\\
$\phi_0,\ \pi_0$&Zero mode of $\phi(x)$ and $\pi(x)$ in normal mode basis\\
$::_a,\ ::_b$&Normal ordering with respect to $A$ or $B$ operators respectively\\
$H, P$&The defining Hamiltonian and corresponding momentum\\
$H\p,\ P\p$&$\df$-transformed $H$ and $P$\\
$H_n$&The $\phi^n$ term in $H^\prime$\\
\hline
Symbol&Description\\
\hline
$f(x)$&The classical kink solution\\
$\df$&Unitary operator that translates $\phi(x)$ by the classical kink solution\\
$g_B(x)$&The kink linearized translation mode\\
$g_k(x)$&Continuum or discrete normal mode\\
$\gamma_i^{mn}$&Coefficient of $\phi_0^m B^{\dag n}\vac_0$ in order $i$ excited state $\ks$\\
$\Gamma_i^{mn}$&Coefficient of $\phi_0^m B^{\dag n}\vac_0$ in order $i$ Schrodinger Equation $(H\p-E)\ks$\\
$V_{ijk}$&Derivative of the potential contracted with various functions\\
$\I(x)$&Contraction factor from Wick's theorem\\
$p$&Momentum\\
$k$&The analog of momentum for normal modes\\
$\kt$&Value of $k$ for the normal mode considered\\
$\omega_k,\ \omega_p$&The frequency corresponding to $k$ or $p$\\
$Q_n$&$n$-loop correction to kink ground state energy\\
$E_n$&$n$-loop correction to excited kink energy\\
\hline
State&Description\\
\hline
$\ks\ (\ks_i)$&Excited kink state as eigenvector of $H\p$ (at order $i$)\\
$\vac\ (\vac_i)$&Kink ground state as eigenvector of $H\p$ (at order $i$)\\
\hline

\end{tabular}
\caption{Summary of Notation}\label{notab}
\end{table}

We now review the formalism introduced in Refs.~\cite{memassa,me2loop} that describes quantum kinks in a 1+1d real scalar field theory with Hamiltonian
\bea
H&=&\int dx \ch(x) \label{hd}\\
\ch(x)&=&\frac{1}{2}:\pi(x)\pi(x):_a+\frac{1}{2}:\partial_x\phi(x)\partial_x\phi(x):_a+\frac{1}{g^2}:V[g\phi(x)]:_a.\nonumber
\eea
The normal-ordering $::_a$ is defined below.  

Consider a kink solution
\beq
\phi(x,t)=f(x) \label{fd}
\eeq
of the classical equations of motion.   We will assume that $V\pp[gf(-\infty)]=V\pp[gf(\infty)]$ and name this quantity $M^2/2$.  Each prime here is a functional derivative with respect to $gf(x)$.

This paper will be entirely in the Schrodinger picture, and so the quantum field $\phi$ only depends on $x$.  One may expand the Schrodinger picture quantum field $\phi(x)$ about its classical solution $\phi(x)=f(x)+\eta(x)$.  In this case $\phi\rightarrow\eta=\phi-f$ could be interpreted as a passive transformation of the fields.  Instead, following \cite{dhn2,rajaraman}, we employ an active transformation of the Hamiltonian and momentum functionals acting on the fields
\beq
H[\phi,\pi]\rightarrow H\p[\phi,\pi]=H[f+\phi,\pi]\hsp
P[\phi,\pi]\rightarrow P\p[\phi,\pi]=P[f+\phi,\pi] \label{act}
.
\eeq
The new observation \cite{mekink} is that this transformation is a unitary equivalence because
\beq
H\p=\df^\dag H\df\hsp P\p=\df^\dag P\df
 \label{hpd}
\eeq
where the displacement operator $\df$ is
\beq
\df={\rm{exp}}\left(-i\int dx f(x)\pi(x)\right). \label{df}
\eeq

It will be necessary to regularize and renormalize the Hamiltonian.  In Eq.~(\ref{hd}) all UV divergences are removed via normal ordering, but this would not be sufficient in theories with fermions or in more dimensions, and so we would like a formalism which may be applied to a general regularized Hamiltonian.   We therefore adopt\footnote{This definition is sufficient to all orders in perturbation theory, however in general to eliminate tadpoles in $H\p$ one must include a correction to $f(x)$ which is exponentially suppressed in the regulator \cite{baiyang}.} (\ref{hpd}) as our definition of $H\p$ and $P\p$ instead of (\ref{act}), as it is well-defined for any regularized Hamiltonian $H$ and agrees with (\ref{act}) when the Hamiltonian is a functional of the unregularized fields.  This approach has the advantage that one regularizes only once.  This is in contrast with the traditional approach in which one separately regularizes $H$ and $H\p$ and so, to remove the regulator at the end of the calculation, one requires a  regulator matching condition that affects the answer \cite{rebhan} but is in general is unknown\footnote{Some matching conditions yield the correct masses in examples at one loop and some do not.  While there are several conjectured principles that determine which are correct \cite{lit,local}, none of these have been derived except in supersymmetric cases.  In the case of theories with a single mass scale, it is often possible to avoid this ambiguity \cite{nastase}.  At one loop the ambiguity can also be avoided \cite{gj}.}.


Unitary equivalence (\ref{hpd}) means that $H$ and $H\p$ have the same eigenvalues, with eigenvectors that are related by $\df$.  This means that we may use whichever is more convenient to calculate any state or energy.  We will see that perturbation theory may be used to calculate vacuum sector states using $H$ and kink sector states using $H\p$.

As $g\sqrt{\hbar}$ is dimensionless\footnote{We set $\hbar=1$.}, we expand $H\p$ in powers of $g$
\bea
H\p&=&\df^\dag H\df=Q_0+\sum_{n=2}^{\infty}H_n\hsp
H_{n(>2)}=\frac{1}{n!}\int dx \V{n}:\phi^n(x):_a\\
H_2&=&\frac{1}{2}\int dx\left[:\pi^2(x):_a+:\left(\partial_x\phi(x)\right)^2:_a+V^{\prime\prime}[gf(x)]:\phi^2(x):_a\right.]\nonumber
\eea
where $Q_0$ is the classical kink mass and $V^{(n)}$ is the $n$th derivative of $g^{n-2}V[g\phi(x)]$ with respect to its argument.  

Consider the classical, linear wave equation corresponding to $H_2$.  The constant frequency\footnote{There are also complex frequency solutions corresponding to quasinormal modes.  In what follows, we will only need our modes to be a basis of the $\delta$-function normalizable functions, or more precisely to satisfy the completeness relation (\ref{comp}).  The real frequency modes alone are sufficient for this goal.  We do not expect the Hamiltonian to mix quasinormal modes with real frequency modes and so quasinormal modes should not contribute to our perturbative calculation of Hamiltonian eigenstates.}
\beq
\omega_k=\sqrt{M^2+k^2} \label{ok}
\eeq
solutions $g_k(x)$ are continuum unbound normal modes, discrete bound normal modes with $0<\omega_k<M$ which we will call shape modes and a zero-mode
\beq
g_B(x)= \frac{f^\prime(x)}{\sqrt{Q_0}}\hsp
\omega_B=0.
\eeq
$k$ is real for continuum modes and imaginary for discrete modes.  The definition (\ref{ok}) of $\omega_k$ fixes the parametrization of $k$ up to a sign.  We will often need to sum over both continuum solutions and shape modes, and so it will be implicit that integrals written $\pin{k}$ also include a sum over the shape modes $\sum_k$.  Similarly, when $k$ represents a shape mode, $2\pi\delta(k-k\p)$ should be understood as $\delta_{kk\p}$.
 
Using the normalization conditions
\beq
\int dx g_{k_1} (x) g^*_{k_2}(x)=2\pi \delta(k_1-k_2),\ 
\int dx |g_{B}(x)|^2=1
\eeq
and conventions
\beq
g_k(-x)=g_k^*(x)=g_{-k}(x),\ \tilde{g}(p)=\int dx g(x) e^{ipx}
\eeq
the completeness relations can be written
\beq
g_B(x)g_B(y)+\pin{k}g_k(x)g^*_{k}(y)=\delta(x-y). \label{comp}
\eeq

Recall that the Schrodinger picture fields $\phi(x)$ and $\pi(x)$ are independent of time.  Therefore, even in the full interacting theory, they may be expanded in any basis of functions.  We will need expansions in terms of plane waves, which diagonalize the free part of $H$
\bea
\phi(x)&=&\pin{p}\left(A^\dag_p+\frac{A_{-p}}{2\omega_p}\right) e^{-ipx}\label{adec}\\
 \pi(x)&=&i\pin{p}\left(\omega_pA^\dag_p-\frac{A_{-p}}{2}\right) e^{-ipx}
\nonumber
\eea
and, following  Ref.~\cite{cahill76}, also normal modes, which diagonalize $H_2$
\bea
\phi(x)&=&\phi_0 g_B(x) +\pin{k}\left(B_k^\dag+\frac{B_{-k}}{2\omega_k}\right) g_k(x)\label{bdec}\\
\pi(x)&=&\pi_0 g_B(x)+i\pin{k}\left(\omega_kB_k^\dag - \frac{B_{-k}}{2}\right) g_k(x).\nonumber
\eea
To simplify later expressions, we have inserted factors of $\sqrt{2\omega}$ into the operators so that $A$ and $A^\dag$, and similarly $B$ and $B^\dag$ are not Hermitian conjugate.  For each decomposition we define a normal ordering.  Plane wave normal ordering $::_a$ places all $A^\dag$ to the left.  Normal mode normal ordering $::_b$ places all $\phi_0$ and $B^\dag$ to the left.  The canonical commutation relations satisfied by $\phi(x)$ and $\pi(x)$ imply
\bea
[A_p,A_q^\dag]&=&2\pi\delta(p-q)\\
{[\phi_0,\pi_0]}&=&i\hsp
[B_{k_1},B^\dag_{k_2}]=2\pi\delta(k_1-k_2).\nonumber
\eea

Our Hamiltonian $H$ is defined in terms of plane wave normal ordering $::_a$.  The unitary transformation (\ref{hpd}) preserves normal ordering \cite{mekink} and so $H\p$ is also plane wave normal-ordered.  Thus $H\p$ is defined in terms of the plane wave operators $A$ and $A^\dag$.  Inserting (\ref{bdec}) into the inverse of (\ref{adec}) one sees that the two sets of operators are related by a linear, Bogoliubov transform.   Using this to express $H\p$ in terms of normal mode operators $B$, $B^\dag$, $\phi_0$ and $\pi_0$ one finds that $H_2$ is a sum of harmonic oscillators with a free particle for the center of mass
\bea
H_2&=&Q_1+\frac{\pi_0^2}{2}+\pin{k}\omega_k B^\dag_k B_k \\
Q_1&=&-\frac{1}{4}\pin{k}\pin{p}\frac{(\omega_p-\omega_k)^2}{\omega_p}\tilde{g}^2_{k}(p)-\frac{1}{4}\pin{p}\omega_p\tilde{g}_{B}(p)\tilde{g}_{B}(p).\nonumber
\eea
Here $Q_1$ is the one-loop kink mass.  The ground state $\vac_0$ of $H_2$ satisfies
\beq
\pi_0\vac_0=B_k\vac_0=0 \label{v0}
\eeq
and corresponds to the one-loop kink ground state.  The exact spectrum of $H_2$ is obtained  by exciting normal modes with $B^\dag_k$ and boosting with $e^{i\phi_0 k/\sqrt{Q_0}}$.  These correspond to the states of the one-kink sector at one loop.  

More generally, the kink ground state corresponds to the eigenstate $\vac$ of $H\p$.  It may be expanded in powers of $\sqrt{\hbar}$
\beq
\vac=\sum_{i=0}^\infty |0\rangle_{i}. \label{semi}
\eeq
The $n$-loop ground state is this sum truncated at $i=2n-2$.

\section{Excited Kink States}  \label{statsez}

\subsection{The Normal Mode State}

Let $\ks$ be the eigenstate of $H\p$ corresponding to a kink with a single excited continuous or discrete normal mode with $k=\kt$.   Note that $\df\ks$ is the corresponding eigenstate of the defining Hamiltonian $H$.  We will use the semiclassical expansion, in powers of $\sqrt{\hbar}$
\beq
\ks=\sum_{i=0}^\infty \ks_i
\eeq
which we will further decompose in terms of normal mode creation operators acting on the state $\vac_0$ 
\beq
\ks_i=\sum_{m,n=0}^\infty \ks_i^{mn}\hsp
\ks_i^{mn}=Q_0^{-i/2}\pink{n}\gamma_{i(\kt)}^{mn}(k_1\cdots k_n)\phi_0^m\Bd1\cdots\Bd n\vac_0. \label{gameq}
\eeq
To avoid clutter, we will leave the $\kt$-dependence of $\gamma$ implicit from here on.

The normal mode $\ks$ is the eigenstate of $H\p$ which, at leading order in the semiclassical expansion, has coefficients
\beq
\gamma_0^{01}(k_1)=2\pi\delta(k_1-\kt) \label{inc}
\eeq
so that at one loop it is simply the harmonic oscillator eigenstate
\beq
\ks_0=\bdk\vac_0.
\eeq
Recall that this is an exact eigenstate of $H_2$, and so it is the correct starting point for our semiclassical expansion of the corresponding eigenstate of $H\p$.  Note that, using our compact notation in which $k$ runs over both real values for continuum modes and discrete indices for shape modes, if $\kt$ is a discrete shape mode then the right side of (\ref{inc}) should be the Kronecker delta $\delta_{k_1\kt}$.  We will continue to write the Dirac delta, reminding the reader that $2\pi\delta$ is always to be read as a Kronecker delta in the discrete case.

\subsection{Translation Invariance}

We will further impose that $\df\ks$ is translation invariant, or equivalently we will work in its center of mass frame.  This condition is
\beq
P\p\ks=0
\eeq
which implies the recursion relations \cite{me2loop,me2looplett}
\bea
\gamma_{i+1}^{mn}(k_1\cdots k_n)&=&\left.\Delta_{k_n B}\left(\gamma_i^{m,n-1}(k_1\cdots k_{n-1})+\frac{\omega_{k_n}}{m}\gamma_i^{m-2,n-1}(k_1\cdots k_{n-1})\right)
\right. \label{rrs}\\
&&+(n+1)\pin{k\p}\Delta_{-k\p B}\left(\frac{\gamma_i^{m,n+1}(k_1\cdots k_n,k\p)}{2\omega_{k\p}}
-\frac{\gamma_i^{m-2,n+1}(k_1\cdots k_n,k\p)}{2m}\right)\nonumber\\
&&+\frac{\omega_{k_{n-1}}\Delta_{k_{n-1}k_n}}{m}\gamma_i^{m-1,n-2}(k_1\cdots k_{n-2})\nonumber\\
&&+\frac{n}{2m}\pin{k\p}\Delta_{k_n,-k\p}\left(1+\frac{\omega_{k_n}}{\omega_{k\p}}\right)\gamma^{m-1,n}_i(k_1\cdots k_{n-1},k\p)
\nonumber\\
&&\left.-\frac{(n+2)(n+1)}{2m}\int\frac{d^2k\p}{(2\pi)^2}\frac{\Delta_{-k\p_1,-k\p_2}}{2\omega_{k\p_2}} \gamma_i^{m-1,n+2}(k_1\cdots k_{n},k\p_1,k\p_2)
\right.
\nonumber
\eea
at all $m>0$.  Here we have defined the  matrix
\beq
\Delta_{ij}=\int dx g_i(x) g\p_j(x).
\eeq
Before each application of the recursion relations, $\gamma_i^{mn}$ must be symmetrized with respect to its arguments $k_j$ \cite{me2loop}.  

The first recursion gives
\bea
\gamma_1^{11}(k_1)&=&\frac{1}{2}\Delta_{k_1,-\kt}\left(1+\frac{\ok1}{\omega_{\kt}}\right)\\
\gamma_1^{13}(k_1,k_2,k_3)&=&\ok2\Delta_{k_2k_3}2\pi\delta(k_1-\kt)\nonumber\\
\gamma_1^{20}&=&-\frac{1}{4}\Delta_{-\kt B}\nonumber\\
\gamma_1^{22}(k_1,k_2)&=&\frac{\ok2}{2}\Delta_{k_2 B}2\pi\delta(k_1-\kt).\nonumber
\eea
Before proceeding to the second recursion, it is necessary to symmetrize the results of the first recursion
\bea
\gamma_1^{13}(k_1,k_2,k_3)&=&\frac{1}{6}\left[\left(\ok2-\ok3\right)\Delta_{k_2k_3}2\pi\delta(k_1-\kt)+\left(\ok1-\ok3\right)\Delta_{k_1k_3}2\pi\delta(k_2-\kt)\right.\nonumber\\
&&\left.+\left(\ok1-\ok2\right)\Delta_{k_1k_2}2\pi\delta(k_3-\kt)
\right]\nonumber\\
\gamma_1^{22}(k_1,k_2)&=&\frac{1}{4}\left[\ok2\Delta_{k_2 B}2\pi\delta(k_1-\kt)+\ok1\Delta_{k_1 B}2\pi\delta(k_2-\kt)\right] \label{g1}
.
\eea

Let us pause to interpret the divergences in these terms.  In the Sine-Gordon model, and we suspect more generally, $\Delta_{k_1k_2}$ contains a summand equal to $-ik_1 2\pi\delta(k_1+k_2)$.  Therefore $\gamma_1^{11}(k_1)$ will have a $\delta(k_1-\kt)$ term.  One can see that with repeated recursions this is part of an $\rm{exp}\left(-i \kt\phi_0/\sqrt{Q_0}\right)\vac_0$ factor of $\ks$.   This term has a simple interpretation.  The condition that $P\p$ annihilates $\ks$, implies that we are working in the center of mass frame of the excited kink.  The operator $B^\dag_\kt$ increases the center of mass momentum by roughly $\kt$ units, and this exponential term compensates with an opposing bulk motion of the kink.  As $\kt/\sqrt{Q_0}$ is of order $g$, this bulk motion is slow, reflecting the fact that the kink is nonperturbatively heavy.  

On the other hand the $\delta(k_1-\kt)$ appearing in $\gamma_1^{13}$ and $\gamma_1^{22}$ reflects the fact that these terms are part of $\bdk\vac_1$.  In other words, they should be interpreted as corrections $
\vac_1$ to the kink ground state $\vac$.    The bare normal mode $\bdk$ is then excited in this dressed ground state.  In this sense, these terms are not caused by the excitation of the normal mode.  To develop a theory of kink scattering, it would be desirable to introduce a suitable LSZ reduction formula.  We suspect that this would eliminate the contributions of such terms to the S-matrix elements in which an asymptotic state is an excited kink $\ks$.

\subsection{Finding Hamiltonian Eigenstates}

The $\gamma_i^{0n}$ are not fixed by translation invariance \cite{me2loop}.  We will now find them using old-fashioned perturbation theory.

In analogy with $\gamma_i^{mn}(k_1\cdots k_n)$, which consists of the $i$th order coefficients of $\ks$ in a basis of the Fock space, we introduce $\Gamma_i^{mn}(k_1\cdots k_n)$ consisting of $i$th order coefficients of $(H\p-E)\ks$.  More precisely, $\Gamma$ is a solution of
\beq
\sum_{j=0}^i \left(H_{i+2-j}-E_{\frac{i-j}{2}+1}\right)\ks_j=Q_0^{-i/2}\sum_{mn} \pink{n} \Gamma_i^{mn}(k_1\cdots k_n)\phi_0^m B_{k_1}^\dag\cdots B_{k_n}^\dag\vac_0.\label{gdef}
\eeq
The $\Gamma$ matrices are clearly functions of the $\gamma$ matrices, as these determine the state $\ks$ via (\ref{gameq}).  

The state $\ks$ is defined to be an eigenvector of $H\p$.  We will refer to the corresponding eigenvalue equation
\beq
(H\p-E)\ks=0\hsp E=\sum_i E_i
\eeq
as the Schrodinger Equation.  Here $E_i$ is the $i$th correction to the energy of $\ks$.  A sufficient condition for a solution is
\beq
\Gamma_i^{mn}(k_1\cdots k_n)=0. \label{gcon}
\eeq
If one symmetrizes this condition over permutations of the arguments $k_j$, then it is also a necessary condition.  

As the $\Gamma$ are functions of the $\gamma$, this condition can be solved for $\gamma$.  We already used translation-invariance to find $\gamma_i^{mn}$ at $m>0$ and so now we need only solve for $\gamma_i^{0n}$.

The leading order is $i=0$.  Recall that
\beq
H_2-Q_1=\frac{\pi_0^2}{2}+\pin{k} \omega_k B_k^\dag B_k \label{libh}
\eeq
and so
\beq
\left(H_2-Q_1\right)\ks_0=\omega_\kt\ks_0.
\eeq
Therefore at leading order (\ref{gdef}) is
\beq
\left(\omega_\kt+Q_1-E_1\right)\ks_0=\sum_{mn} \pink{n} \Gamma_0^{mn}(k_1\cdots k_n)\phi_0^m B_{k_1}^\dag\cdots B_{k_n}^\dag\vac_0.
\eeq
The condition $\Gamma_0=0$ implies
\beq
E_1=Q_1+\omega_\kt. \label{e1}
\eeq
This is not a big surprise, it is just the statement that at leading order the mass $E_1$ of a kink with an excited normal mode is greater than the ground state kink mass $Q_1$ by $\omega_\kt$.

The next order is $i=1$, where we find
\beq
H_3\ks_0+(H_2-E_1)\ks_1=Q_0^{-1/2}\sum_{mn} \pink{n} \Gamma_1^{mn}(k_1\cdots k_n)\phi_0^m B_{k_1}^\dag\cdots B_{k_n}^\dag\vac_0.
\eeq
Using (\ref{e1}) we see that
\beq
H_2-E_1=-\omega_\kt+\frac{\pi_0^2}{2}+\pin{k} \omega_k B_k^\dag B_k. \label{he1}
\eeq
Recall that
\bea
H_3&=&\frac{1}{6}\int dx \V3:\phi^3(x):_a\\
&=&\frac{1}{6}\int dx \V3:\phi^3(x):_b+\frac{1}{2}\int dx \V3\phi(x) \I(x).\nonumber
\eea
In the second line we have used Wick's theorem \cite{mewick} where the contraction factor $\I(x)$ is defined by
\beq
\partial_x \I(x)=\pin{k}\frac{1}{2\omega_k}\partial_x\left|g_{k}(x)\right|^2 \label{di}
\eeq
with the boundary condition fixed so that $\I(x)$ vanishes asymptotically. 

Let us calculate the entries $\Gamma_1^{0n}$ one at a time.  Introducing the notation
\beq
V_{\I\stackrel{m}{\cdots}\I,\alpha_1\cdots\alpha_n}=\int dx V^{(2m+n)}[gf(x)]\I^m(x) g_{\alpha_1}(x)\cdots g_{\alpha_n(x)}
\eeq
there are three contributions to $\Gamma_1^{00}$
\bea
H_3\ks_0&\supset& \frac{V_{\I -\kt}}{4\omega_{\kt}}\vac_0\\
\frac{\pi_0^2}{2}\ks_1&\supset&\frac{\pi_0^2}{2}Q_0^{-1/2}\gamma_{1}^{20}\phi_0^2\vac_0=\frac{1}{4\sqrt{Q_0}}\Delta_{-\kt B}\vac_0\nonumber\\
-\omega_\kt\ks_1&\supset&-\frac{\omega_\kt}{\sqrt{Q_0}}\gamma_{1}^{00}\vac_0.
\eea
These lead to
\beq
\Gamma_1^{00}= \frac{\sqrt{Q_0}V_{\I -\kt}}{4\omega_{\kt}}+\frac{\Delta_{-\kt B}}{4}-\omega_\kt\gamma_{1}^{00}.
\eeq
Schrodinger's equation $\Gamma=0$ then yields
\beq
\gamma_1^{00}= \frac{\sqrt{Q_0}V_{\I -\kt}}{4\omega^2_{\kt}}+\frac{\Delta_{-\kt B}}{4\omega_\kt}.
\eeq

The contributions to $\Gamma_1^{02}$ are similar, but there is also a contribution from $:\phi^3:_b\ks_0$, or more precisely from terms of the form $B^\dag B^\dag B\ks_0$.  Altogether we find
\bea
H_3\ks_0&\supset& \frac{1}{2}\pin{k} V_{\I k} B^\dag_{k}\bdk\vac_0+\frac{1}{2}\pink{2} \frac{V_{-\kt k_1k_2}}{2\omega_\kt}\Bd 1\Bd2\vac_0
\\
\frac{\pi_0^2}{2}\ks_1&\supset&\frac{\pi_0^2}{2}Q_0^{-1/2}\pink{2}\gamma_{1}^{22}(k_1,k_2)\phi_0^2\Bd{1}\Bd{2}\vac_0=-\frac{1}{2\sqrt{Q_0}}\pin{k}\ok{}\Delta_{k B}\Bd{}\bdk\vac_0\nonumber
\eea
and
\beq
\left(-\omega_\kt+\pin{k} \omega_k B_k^\dag B_k\right)\ks_1\supset\frac{1}{\sqrt{Q_0}}\pink{2}\left(\ok1+\ok2-\omega_\kt\right)\gamma_{1}^{02}(k_1,k_2)\Bd{1}\Bd{2}\vac_0.
\eeq
Adding these contributions we find
\beq
\Gamma_1^{02}=  \frac{2\pi\delta(k_2-\kt)}{2}\left(\sqrt{Q_0}V_{\I  k_1}-\ok1\Delta_{k_1 B}\right)+\frac{\sqrt{Q_0}}{2}\frac{V_{-\kt k_1 k_2}}{2\omega_\kt}+\left(\ok1+\ok2-\omega_\kt\right)\gamma_1^{02}(k_1,k_2).
\eeq
Schrodinger's equation $\Gamma=0$ then yields
\beq
\gamma_1^{02}(k_1,k_2)= \frac{2\pi\delta(k_2-\kt)}{2}\left(\Delta_{k_1 B}-\sqrt{Q_0}\frac{V_{\I  k_1}}{\ok1}\right)+\frac{\sqrt{Q_0}V_{-\kt k_1 k_2}}{4\omega_\kt\left(\omega_\kt-\ok1-\ok2\right)}.
\eeq
As we will insert this into the recursion relation (\ref{rrs}) later, we will need its symmetrized form
\bea
\gamma_1^{02}(k_1,k_2)&=& \frac{2\pi\delta(k_2-\kt)}{4}\left(\Delta_{k_1 B}-\sqrt{Q_0}\frac{V_{\I  k_1}}{\ok1}\right)\\
&&+\frac{2\pi\delta(k_1-\kt)}{4}\left(\Delta_{k_2 B}-\sqrt{Q_0}\frac{V_{\I  k_2}}{\ok2}\right)+\frac{\sqrt{Q_0}V_{-\kt k_1 k_2}}{4\omega_\kt\left(\omega_\kt-\ok1-\ok2\right)}.\nonumber
\eea

Finally, we will compute $\Gamma_1^{04}$.  As $\gamma_1^{24}=0$ there are only two contributions
\beq
H_3\ks_0\supset \frac{1}{6}\pink{3} V_{k_1k_2k_3}\Bd 1\Bd2\Bd3\bdk\vac_0
\eeq
and
\beq
\left(-\omega_\kt+\pin{k} \omega_k B_k^\dag B_k\right)\ks_1\supset\frac{1}{\sqrt{Q_0}}\pink{4}\left(-\omega_\kt+\sum_{j=1}^4 \ok{j}\right)\gamma_{1}^{04}(k_1\cdots k_4)\Bd{1}\cdots \Bd{4}\vac_0
\eeq
leading to
\beq
\Gamma_1^{04}=  \frac{2\pi\delta(k_4-\kt)}{6}V_{k_1k_2k_3}+\left(-\omega_\kt+\sum_{j=1}^4 \ok{j}\right)\gamma_1^{04}(k_1\cdots k_4).
\eeq
Thus the last matrix element at order $i=1$ is
\beq
\gamma_1^{04}(k_1\cdots k_4)= -\frac{\sqrt{Q_0}V_{k_1 k_2 k_3}}{6\sum_{j=1}^3 \ok{j}}2\pi\delta(k_4-\kt).
\eeq

This completes our determination of $\gamma_1^{mn}$ and so of the leading correction $\ks_1$ to the excited kink state $\ks$.

\section{Mass Shifts} \label{masssez}

In this section we will calculate the leading order correction to the masses of the normal modes.  More precisely, $E_2$ will be the two-loop correction to the energy of the excited kink.  Subtracting $Q_2$, the two-loop correction to the ground state energy found in Ref.~\cite{me2looplett}, one obtains $E_2-Q_2$, the two-loop correction to the energy required to excite the kink normal mode.

\subsection{The Next Order Schrodinger Equation}

The leading order energy correction is $E_2$, which can be computed from the $i=2$ Schrodinger equation
\beq
(H_4-E_2)\ks_0+H_3\ks_1+(H_2-E_1)\ks_2=0. \label{s2}
\eeq
As $\ks_0=\bdk\vac_0$, the energy $E_2$ is fixed by terms that are proportional to $\bdk\vac_0$.  

More precisely, we need only calculate $\Gamma_2^{01}$.   In Sec.~\ref{statsez} we fixed $\ks_0$ and found $\ks_1$.  The only terms in $\ks_2$ that contribute to $\Gamma_2^{01}$ are $\gamma_2^{01}$ and $\gamma_2^{21}$, the first via the $-\omega_\kt+\pin{k} \omega_k B_k^\dag B_k$ term in $H_2$ and the second via the $\pi_0^2/2$ term.  

At second order, the only $m>0$ contribution to the energy arises from $\gamma_2^{21}$ as the $\pi_0^2$ maps it to the initial state $m=0$, $n=1$.  Using the recursion relation (\ref{rrs}) this is given by
\bea
\gamma_2^{21}(k_1)&=&\left.\Delta_{k_1 B}\left(\gamma_1^{20}+\frac{\omega_{k_1}}{2}\gamma_1^{00}\right)
\right.+2\pin{k\p}\Delta_{-k\p B}\left(\frac{\gamma_1^{22}(k_1,k\p)}{2\omega_{k\p}}
-\frac{\gamma_1^{02}(k_1,k\p)}{4}\right)\\
&&+\frac{1}{4}\pin{k\p}\Delta_{k_1,-k\p}\left(1+\frac{\omega_{k_1}}{\omega_{k\p}}\right)\gamma^{11}_1(k\p)
-\frac{3}{2}\int\frac{d^2k\p}{(2\pi)^2}\frac{\Delta_{-k\p_1,-k\p_2}}{2\omega_{k\p_2}} \gamma_1^{13}(k_1,k\p_1,k\p_2).\nonumber
\eea
Inserting the coefficients $\gamma_0$ and $\gamma_1$ found in Sec.~\ref{statsez} this becomes
\bea
\gamma_2^{21}(k_1)&=&
2\pi\delta(k_1-\kt)\left[\pin{k\p}\Delta_{-k\p B}\Delta_{k\p B}\left(\frac{1}{4}-\frac{1}{8}\right)+\frac{1}{8}\Delta_{-k\p B}\frac{\sqrt{Q_0}V_{\I  k\p}}{\okp{}}\right.\\
&&\left.
+\frac{1}{8}\int\frac{d^2k\p}{(2\pi)^2}\left(1-\frac{\okp1}{\okp2}\right)\Delta_{k\p_1k\p_2}\Delta_{-k\p_1,-k\p_2}\right]\nonumber\\
&&+\left[-\left(\frac{1}{4}+\frac{1}{8}\right)+\left(\frac{1}{8}+\frac{1}{4}\right)\frac{\ok1}{\omega_{\kt}}\right]\Delta_{k_1 B}\Delta_{-\kt B}\nonumber\\
&&+\frac{\sqrt{Q_0}}{8\omega_\kt}\left(\omega_{k_1}\Delta_{k_1 B}\frac{V_{\I -\kt}}{\omega_\kt}+\omega_{\kt}\Delta_{-\kt B}\frac{V_{\I  k_1}}{\ok1}\right)-\frac{1}{2}\pin{k\p}\Delta_{-k\p B}
\frac{\sqrt{Q_0}V_{-\kt k_1 k\p}}{4\omega_\kt\left(\omega_\kt-\ok1-\okp{}\right)}\nonumber\\
&&-\frac{1}{8}\pin{k\p}\left[\left(1+\frac{\omega_{k_1}}{\omega_{\kt}}\right)(1-1)+\left(\frac{\ok1}{\okp{}}+\frac{\okp{}}{\omega_{\kt}}
\right)(1+1)\right]\Delta_{-\kt,-k\p}\Delta_{k_1k\p}.\nonumber
\eea
Simplifying slightly this is
\bea
\gamma_2^{21}(k_1)&=&
2\pi\delta(k_1-\kt)\left[\pin{k\p}\frac{\Delta_{-k\p B}}{8}\left(\Delta_{k\p B}+\frac{\sqrt{Q_0}V_{\I  k\p}}{\okp{}}\right)\right.\\
&&\left.
-\frac{1}{16}\int\frac{d^2k\p}{(2\pi)^2}\frac{\left(\okp1-\okp2\right)^2}{\okp1\okp2}\Delta_{k\p_1k\p_2}\Delta_{-k\p_1,-k\p_2}\right]\nonumber\\
&&+\frac{3}{8}\left(-1+\frac{\ok1}{\omega_{\kt}}\right)\Delta_{k_1 B}\Delta_{-\kt B}-\frac{1}{4}\pin{k\p}\left(\frac{\ok1}{\okp{}}+\frac{\okp{}}{\omega_{\kt}}
\right)\Delta_{-\kt,-k\p}\Delta_{k_1k\p}\nonumber\\
&&+\frac{\sqrt{Q_0}}{8\omega_\kt}\left(\omega_{k_1}\Delta_{k_1 B}\frac{V_{\I -\kt}}{\omega_\kt}+\omega_{\kt}\Delta_{-\kt B}\frac{V_{\I  k_1}}{\ok1}\right)-\frac{1}{2}\pin{k\p}\Delta_{-k\p B}
\frac{\sqrt{Q_0}V_{-\kt k_1 k\p}}{4\omega_\kt\left(\omega_\kt-\ok1-\okp{}\right)}.\nonumber
\eea

Now we will compute the various contributions to $\Gamma_2^{01}$.  Let us begin with the contributions to $(H_2-E_1)\ks_2$ in (\ref{s2}).  The operator is given in (\ref{he1}).  The contribution from $\gamma_2^{21}$ arises from
\beq
\frac{\pi_0^2}{2}\frac{1}{Q_0}\pink{1}\gamma_2^{21}(k_1)\phi_0^2\Bd{1}\vac_0=-\frac{1}{Q_0}\pink{1}\gamma_2^{21}(k_1)\Bd{1}\vac_0.
\eeq
The contribution of $\gamma_2^{01}$ is
\beq
\left(-\omega_\kt+\pin{k} \omega_k B_k^\dag B_k\right)\frac{1}{Q_0}\pink{1}\gamma_2^{01}(k_1)\Bd{1}\vac_0=\frac{1}{Q_0}\pink{1}\left(\ok1-\omega_\kt\right)\gamma_2^{01}(k_1)\Bd{1}\vac_0.
\eeq
The contribution to the energy arises from $k_1=\kt$ but in that case the $\ok1-\omega_\kt$ vanishes and so this term does not contribute.  This is an important consistency check, as $\gamma_2^{01}(\kt)$ can be absorbed into the arbitrary normalization of $\gamma_0^{01}(\kt)$ and this choice should not affect an observable quantity like the energy.

There are three contributions from $H_3\ks_1$.  The first is
\bea
H_3\ks_1^{00}&=&\frac{1}{\sqrt{Q_0}}\gamma_1^{00}H_3\vac_0\supset\frac{1}{2\sqrt{Q_0}}\gamma_1^{00}\pink{1}V_{\I  k_1}\Bd1\vac_0\\
&=&\frac{1}{8}\left(\frac{V_{\I  -\kt}}{\omega^2_{\kt}}+\frac{\Delta_{-\kt B}}{\omega_\kt\sqrt{Q_0}}\right)\pink{1}V_{\I k_1}\Bd1\vac_0.\nonumber
\eea
The second is
\bea
H_3\ks_1^{02}&=&\frac{1}{\sqrt{Q_0}}\pink{2}\gamma_1^{02}(k_1,k_2)H_3\Bd1\Bd2\vac_0\nonumber\\
&\supset&\frac{1}{\sqrt{Q_0}}\pink{2}\gamma_1^{02}(k_1,k_2)\left[\frac{6}{6}\pinkp{3}V_{-k\p_1-k\p_2k\p_3}\frac{2\pi\delta(k_1-k\p_1)}{2\okp1}\frac{2\pi\delta(k_2-k\p_2)}{2\okp2}\Bd3\right.\nonumber\\
&&\left.+\frac{2}{2}\pinkp{1}V_{\I -k\p_1}\frac{2\pi\delta(k_1-k\p_1)}{2\okp1}\Bd2
\right]\vac_0\nonumber\\
&=&\frac{1}{\sqrt{Q_0}}\pink{1}\left[\pinkp{2}\frac{\gamma_1^{02}(k\p_1,k\p_2)V_{-k\p_1-k\p_2k_1}}{4\okp1\okp2}
+\pinkp{1}\frac{\gamma_1^{02}(k\p_1,k_1)V_{\I -k\p_1}}{2\okp1}\right]\Bd1\vac_0\nonumber\\
&=&\frac{1}{\sqrt{Q_0}}\pink{1}\left[\pinkp{2}\frac{\sqrt{Q_0}V_{-\kt k\p_1 k\p_2}V_{-k\p_1-k\p_2k_1}}{16\omega_\kt\okp1\okp2\left(\omega_\kt-\okp1-\okp2\right)}\right.\nonumber\\
&&\left.+\pinkp{1}\left(\frac{\left(\okp1\Delta_{k\p_1 B}-\sqrt{Q_0}V_{\I  k\p_1}\right)
V_{-k\p_1-\kt k_1}}{8\okp1^2\omega_\kt}+\frac{\sqrt{Q_0}V_{-\kt k\p_1 k_1}V_{\I -k\p_1}}{8\omega_\kt\okp1\left(\omega_\kt-\okp1-\ok1\right)}\right)\right.\nonumber\\
&&+\left.
\frac{ \left(\ok1\Delta_{k_1 B}-\sqrt{Q_0}V_{\I  k_1}\right)V_{\I -\kt}}{8\omega_\kt\ok1}\right.\nonumber\\
&&\left.
+2\pi\delta(k_1-\kt)\pinkp{1}\frac{\left(\okp1\Delta_{k\p_1 B}-\sqrt{Q_0}V_{\I  k\p_1}\right)
V_{\I -k\p_1}}{8\okp1^2}
\right]\Bd1\vac_0.\nonumber
\eea
The third contribution is
\bea
H_3\ks_1^{04}&=&\frac{1}{\sqrt{Q_0}}\pink{4}\gamma_1^{04}(k_1\cdots k_4)H_3\Bd1\cdots\Bd4\vac_0\label{ter}\\
&\supset&\frac{1}{\sqrt{Q_0}}\pink{4}\left[-\frac{\sqrt{Q_0}V_{k_1 k_2 k_3}}{6\sum_{j=1}^3 \ok{j}}2\pi\delta(k_4-\kt)\right]\frac{1}{6}\pinkp{3}V_{-k\p_1-k\p_2-k\p_3}\nonumber\\
&&\left.\times\left(6\frac{2\pi\delta(k_1-k\p_1)}{2\okp1}\frac{2\pi\delta(k_2-k\p_2)}{2\okp2}\frac{2\pi\delta(k_3-k\p_3)}{2\okp3}\Bd4\right.\right.\nonumber\\
&&+\left.18\frac{2\pi\delta(k_4-k\p_3)}{2\okp1}\frac{2\pi\delta(k_2-k\p_1)}{2\okp2}\frac{2\pi\delta(k_3-k\p_2)}{2\okp3}\Bd1
\right]\vac_0\nonumber\\
&&=-\frac{1}{48}\pink{1}\left[3\pinkp{2}\frac{V_{k_1k\p_1k\p_2}V_{-\kt-k\p_1-k\p_2}}{\omega_\kt\okp1\okp2\left(\ok1+\okp1+\okp2\right)}\right.\nonumber\\
&&\left.+2\pi\delta(k_1-\kt)\pinkp{3}\frac{V_{k\p_1k\p_2k\p_3}V_{-k\p_1-k\p_2-k\p_3}}{\okp1\okp2\okp3\left(\okp1+\okp2+\okp3\right)}
\right]\Bd1\vac_0.\nonumber
\eea

The last contribution to $\Gamma_2^{01}$ is from $(H_4-E_2)\ks_0^{01}$.  This is easily evaluated using Wick's theorem \cite{mewick}
\beq
H_4\ks_0^{01}\supset \left(\frac{V_{\I \I}}{8}+\pinkp{2}\frac{V_{\I k_1 k_2}}{2}B^\dag_{k\p_1}\frac{B_{-k\p_2}}{2\okp2}\right)B^\dag_{\kt}\vac_0
=\frac{V_{\I \I}}{8}\ks_0^{01}+\pink{1}\frac{V_{\I k_1 -\kt}}{4\omega_\kt}\Bd1\vac_0.
\eeq
Therefore
\beq
(H_4-E_2)\ks_0\supset\pink{1}\left[\left(\frac{V_{\I\I}}{8}-E_2\right)2\pi\delta(k_1-\kt)+\frac{V_{\I k_1 -\kt}}{4\omega_\kt}
\right]\Bd1\vac_0.
\eeq

Summing all of these contributions, one finds
\beq
0=\frac{\Gamma_2^{01}(k_1)}{Q_0}=(Q_2-E_2) 2\pi\delta(k_1-\kt)+\mu(k_1) \label{g2}
\eeq
for some function $\mu(k_1)$.   $Q_2$ is the two-loop kink ground state energy found in Ref.~\cite{me2loop} and repeated here in Eq.~(\ref{q2}).  

While $\mu(k_1)$ is somewhat lengthy, only the case $k_1=\kt$ is relevant to the discussion of mass corrections\footnote{The fact that $\mu(k_1)$ vanishes at $k_1\neq \kt$ fixes $\gamma_2^{01}(k_1)$.}
\bea
\mu(\kt)&=&\pinkp{2}\frac{\left(\okp1+\okp2\right)V_{-\kt k\p_1 k\p_2}V_{-k\p_1-k\p_2\kt}}{8\omega_\kt\okp1\okp2\left(\omega_\kt^2-\left(\okp1+\okp2\right)^2\right)}-\pin{k\p}\frac{V_{-\kt k\p \kt}V_{\I -k\p}}{4\omega_\kt\okp{}^2}+\frac{V_{\I \kt -\kt}}{4\omega_\kt}\nonumber\\
&&
+\frac{1}{4Q_0}\pin{k\p}\left(\frac{\omega_\kt}{\okp{}}+\frac{\okp{}}{\omega_{\kt}}
\right)\Delta_{-\kt-k\p}\Delta_{\kt k\p}.\label{muk}
\eea

We note in passing that the two $\I$ terms have an interesting property.  If they are integrated over $\kt$, they produce exactly twice the first two terms in the $Q_2$.  This is reminiscent of the quantum harmonic oscillator, where the ground state energy is $\omega/2$ and each excited state produces an additional $\omega$, which is twice the ground state contribution.  Thus in a free theory this relationship between the kink ground state energy $Q_2$ and the normal mode excitation energy $\mu(\kt)$ would be expected.  But why does it appear here?  The reason is that if we normal mode normal order the kink Hamiltonian $H\p$, then Wick's theorem implies that the interaction terms $H_3$ and $H_4$ contribute to the linear and quadratic parts of the normal mode normal-ordered $H\p$, with a contribution given by folding the $\I$ factor from Wick's theorem into the corresponding potential $V^{(3)}$ or $V^{(4)}$.  These new contributions to the free part of the Hamiltonian shift the oscillator frequencies by quantities proportional to various $V_\I$, but suppressed by a power of the coupling as they arose from $H_3$ or $H_4$.  Then, since the leading contribution to the (kink) ground state energy is half the integral of the normal mode frequencies, it is shifted by the integral of half of this frequency shift, while each excitation of a normal mode increases the energy by the frequency.

\subsection{Continuum Modes}

If $\kt$ is a continuum mode then the term with the Dirac $\delta$ in (\ref{g2}) is infinite and so,   if $\mu(\kt)$ is finite, must vanish separately.  This implies $E_2=Q_2$ for all continuum modes $\kt$ with $\mu(\kt)$ finite.   This is intuitive, the continuum modes are nonnormalizable and they only have finite overlap with the kink.  Therefore the kink cannot shift their energy.  The two-loop correction to the energy needed to excite the kink ground state to a normal mode state is $E_2-Q_2=0$.   Of course the one-loop correction $E_1-Q_1=\ok{}$ we have already seen is nonzero.

Is $\mu(\kt)$ finite?  Divergences may only arise from divergences in $\Delta$ or $V$ or from the infinite integrals over continuum modes $k$.  If the potential $V$ is smooth then divergences in the $V$ symbols will not arise.  However $\Delta$ has a divergence arising from the fact that continuum modes tend to plane waves far from a localized kink
\beq
\Delta_{k_1k_2}\supset i\pi (k_2-k_1)\delta(k_1+k_2).
\eeq
Via $\gamma_2^{21}(k_1)$, this contributes
\beq
\mu(k_1)\supset \frac{\kt^2}{2Q_0}2\pi\delta(\kt-k_1).
\eeq
If there are no other divergences in $\mu$ then, setting to zero the coefficient of $\delta(\kt-k_1)$ in (\ref{g2}), we find
\beq
E_2=Q_2+\frac{\kt^2}{2Q_0}.
\eeq
In other words, the two-loop correction to the mass of a kink excited by a normal mode with $k=\kt$ is just the corresponding nonrelativistic kinetic energy.  

The appearance of the nonrelativistic kinetic energy may be surprising as we are in the kink center of mass frame.  However this is actually the nonrelativistic energy resulting from the fact that, as described beneath Eq.~(\ref{g1}), in order to keep a total momentum of zero the nonrelativistic kink has a bulk motion which compensates that of the relativistic normal mode.  Due to the mass difference, the kinetic energy of the normal mode affects the total energy at one loop while the kinetic energy of the bulk, which has an equal and opposite momentum, enters only at two loops.

Now let us consider potential divergences in the $k$ integrals.  The corresponding eigenfunctions tend to plane waves $e^{ikx}$ far from the kink, up to a phase shift.  In cases such as the Sine-Gordon and $\phi^4$ models the $\Delta$ and $V_{ijk}$ tend exponentially to zero in the sum of their indices, as the theories are gapped.  Therefore the only divergence may arise from an infinite domain of integration in which the sum of the indices is within a fixed distance of zero.  This requires a double integral, with $k\p_1\sim-k\p_2$, and so divergences may only arise in the first term of (\ref{muk}).  

At the large $k\p$ on which these divergences are supported, $g_{k\p}(x)\sim e^{ik\p x}$.  The divergence is also supported at large $x$, where $V^{(3)}$ tends to a constant, which on each side of the kink is just the third derivative of the potential supported on the corresponding vacuum.  Let us say for simplicity that these two third derivatives have the same value, $W$, up to a sign.  This is the case in the $\phi^4$ model, whereas in the Sine-Gordon model the third derivatives vanish at the vacua so $W=0$.  We have argued that, up to finite terms
\beq
V_{-\kt k\p_1 k\p_2}\sim V_{-k\p_1-k\p_2\kt}\sim  W \int dx e^{i(k\p_1+k\p_2-\kt)x}=W 2\pi\delta(\kt-k\p_1-k\p_2). \label{vdiv}
\eeq

Thus there are two $\delta$ functions in the third integrand of (\ref{muk}).  The first may be used to do one of the integrals, but then the other is a genuine $\delta$ function divergence
\beq
\mu(k_1)\sim \frac{W^2 2\pi\delta(k_1-\kt)}{8}\pin{k\p}\frac{\okp{}+\omega_{\kt-\okp{}}}{\okp{}\omega_{\kt-\okp{}}\left(\omega_\kt^2-\left(\okp{}+\omega_{\kt-\okp{}}\right)^2\right)}.
\eeq 
It combines with the $Q_2$ term to shift the energy $E_2$ by
\beq
\frac{W^2}{8}\pin{k\p}\frac{\okp{}+\omega_{\kt-\okp{}}}{\okp{}\omega_{\kt-\okp{}}\left(\omega_\kt^2-\left(\okp{}+\omega_{\kt-\okp{}}\right)^2\right)}
\eeq
which just yields the usual one-loop correction to the mass of the plane wave in the absence of the kink.  It shifts the mass of the normal mode.


\subsection{Shape Modes}

In the case of shape modes, one recalls that the $2\pi\delta(k_1-\kt)$ in $\gamma_0^{01}(k_1)$ is to be replaced by the Kronecker delta $\delta_{k_1\kt}$.  Thus (\ref{g2}) evaluated at $k_1=\kt$ is finite
\beq
\frac{\Gamma_2^{01}(\kt)}{Q_0}=(Q_2-E_2) +\mu(\kt).
\eeq

The Schrodinger equation $\Gamma=0$ then yields
\beq
E_2=Q_2+\mu(\kt). \label{bm}
\eeq
Again $\mu(\kt)$ is given by (\ref{muk}).  However the divergence (\ref{vdiv}) does not arise because $g_\kt(x)$ is a bound state of the potential and so falls to zero at large $x$, exponentially in the case of the Sine-Gordon or $\phi^4$ models.  This absence of divergences is fortunate as a divergent $\mu(\kt)$ would in this case have led to a divergent $E_2$ as a result of (\ref{bm}).

\section{A Diagrammatic Approach} \label{dessinsez}

\subsection{The Kink Ground State}

The two-loop energy of the kink ground state is \cite{me2loop}
\bea
Q_2&=&\frac{V_{\I \I}}{8}-\frac{1}{8}\pin{k\p}\frac{\left|V_{\I  k\p}\right|^2}{\okp{}^2}
-\frac{1}{48}\pinkp{3} \frac{\left|V_{k\p_1k\p_2k\p_3}\right|^2}{\omega_{k\p_1}\omega_{k\p_2}\omega_{k\p_3}\left(\omega_{k\p_1}+\omega_{k\p_2}+\omega_{k\p_3}\right)}\label{q2}\\
&&  +\frac{1}{16Q_0}\pinkp{2}\frac{\left|\left(\omega_{k_1\p}-\omega_{k_2\p}\right)\Delta_{k_1\p k_2\p}\right|^2}{\omega_{k\p_1}\omega_{k\p_2}}  -\frac{1}{8Q_0}\pin{k\p}\left|\Delta_{k\p B}\right|^2. 
\nonumber
\eea
Recalling from Refs.~\cite{me2stato} that
\beq
 V_{BBk}=-\frac{\omega_k^2}{\sqrt{Q_0}}\Delta_{kB}\hsp
 V_{Bk_1k_2}=\frac{\omega_{k_2}^2-\omega_{k_1}^2}{\sqrt{Q_0}}\Delta_{k_1 k_2} \label{vid}
 \eeq
 the last two terms may be reexpressed in terms of $|V_{Bk\p_1k\p_2}|^2$ and $ |V_{BBk\p}|^2$ respectively
\bea
Q_2&=&\frac{V_{\I \I}}{8}-\frac{1}{8}\pin{k\p}\frac{\left|V_{\I  k\p}\right|^2}{\okp{}^2}
-\frac{1}{48}\pinkp{3} \frac{\left|V_{k\p_1k\p_2k\p_3}\right|^2}{\omega_{k\p_1}\omega_{k\p_2}\omega_{k\p_3}\left(\omega_{k\p_1}+\omega_{k\p_2}+\omega_{k\p_3}\right)}\\
&&  +\frac{1}{16}\pinkp{2}\frac{\left|V_{Bk\p_1k\p_2}\right|^2}{\omega_{k\p_1}\omega_{k\p_2}\left(\okp1+\okp2\right)^2}  -\frac{1}{8}\pin{k\p}\frac{\left|V_{BBk\p}\right|^2}{\okp{}^4}. 
\nonumber
\eea

\begin{figure}
\setlength{\unitlength}{.7cm}
\begin{picture}(6,4)(-7,-1)
\put(-2,1){\circle{2}}
\put(-4,1){\circle{2}}
\put(-3.01,1){\circle*{.1}}

\put(6,1){\circle{2}}
\put(5,1){\vl{-1}0 {.5}}
\put(3,1){\circle{2}}
\put(4.6,1.3){\makebox(0,0){{$k\p$}}}

\qbezier(10,1)(12,3)(14,1)
\put(14,1){\vl{-1}0 2}
\put(12,2){\vector(-1,0) 0}
\put(12,0){\vector(-1,0) 0}
\qbezier(10,1)(12,-1)(14,1)
\put(12,2.4){\makebox(0,0){{$k\p_1$}}}
\put(12,1.4){\makebox(0,0){{$k\p_2$}}}
\put(12,0.4){\makebox(0,0){{$k\p_3$}}}




\end{picture}
\caption{Diagrams corresponding to the first three terms in $Q_2$.  Every vertex is an interaction in $H\p$.  Operator ordering runs to the left.  Each loop gives a factor of $\I$.}
\label{q2fig}
\end{figure}
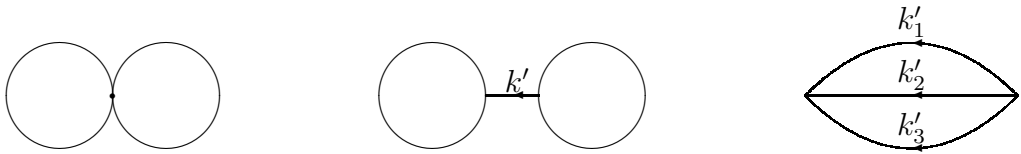

The first three terms in $Q_2$ are easily calculated using the diagrams in Fig.~\ref{q2fig} to represent various contributions to $H\p\vac$.  Operator ordering runs to the left.  Each loop involving a single vertex brings a factor of $\I(x)$ and each $n$-point vertex brings a $V^{(n)}$ which is integrated over $x$ together with the normal modes  $g_k(x)$ arising from the attached lines and loop factors $\I(x)$ from attached loops.  Each internal line corresponding to a normal mode $k$ brings a factor of $1/(2{\ok{}} )$.  In addition, each vertex except for the last brings a factor $\left(\sum_i \omega_i-\sum_j\omega_j\right)^{-1}$ where $i$ runs over all outgoing $k$ and $j$ runs over all incoming $k$.  Symmetry factors are calculated as for Feynman diagrams, for example in the first term each loop may be inverted and the two may be interchanged leading to a symmetry factor of $(1/2)^3$.  In the second each loop may be inverted leading to $(1/2)^2$ while in the third the three propagators may be exchanged leading to $1/6$.

What about the fourth and fifth terms?  Clearly a corresponding diagram may be drawn by taking the third diagram in Fig.~\ref{q2fig} and replacing one or two normal mode lines $k\p$ with a zero-mode line $B$.  However one may choose whether the vertices are to be constructed using $H\p$ or $P\p$.  At higher orders this distinction is important because, for example, in the Sine-Gordon theory $H\p$ has an infinite number of terms whereas in any theory $P\p$ has only one term for each summand in the recursion relation (\ref{rrs}).  Thus there are multiple possible conventions for representing these terms diagrammatically, and the Feynman diagram convention of allowing each vertex to represent an interaction in $H\p$ is not the most economical.  We will leave the development of diagrammatic methods using $P\p$ vertices to future work.

\subsection{Normal Modes}

\begin{figure}
\setlength{\unitlength}{.7cm}
\begin{picture}(6,4)(-7,-1)
\put(-2,1){\circle{2}}
\put(-4,1){\circle{2}}
\put(-3.01,1){\circle*{.1}}
\put(-1,-1){\vl{-1}0 2}
\put(-2.9,-.7){\makebox(0,0){{$\kt$}}}

\put(6,1){\circle{2}}
\put(5,1){\vl{-1}0 {.5}}
\put(3,1){\circle{2}}
\put(4.6,1.3){\makebox(0,0){{$k\p$}}}
\put(7,-1){\vl{-1}0 {2.5}}
\put(4.6,-.7){\makebox(0,0){{$\kt$}}}

\qbezier(10,1)(12,3)(14,1)
\put(14,1){\vl{-1}0 2}
\put(12,2){\vector(-1,0) 0}
\put(12,0){\vector(-1,0) 0}
\qbezier(10,1)(12,-1)(14,1)
\put(12,2.4){\makebox(0,0){{$k\p_1$}}}
\put(12,1.4){\makebox(0,0){{$k\p_2$}}}
\put(12,0.4){\makebox(0,0){{$k\p_3$}}}
\put(14,-1){\vl{-1}0 2}
\put(12.1,-.7){\makebox(0,0){{$\kt$}}}

\end{picture}
\caption{Diagrams corresponding to the first three terms in the $Q_2 2\pi\delta(k_1-\kt)$ contribution to $E_2$.  The $k_1=\kt$ line is disconnected from the diagram.  Therefore these are just contributions to the ground state energy $Q_2$, and so they do not contribute to the energy $E_2-Q_2$ needed to excite a normal mode in the kink background.}
\label{discfig}
\end{figure}
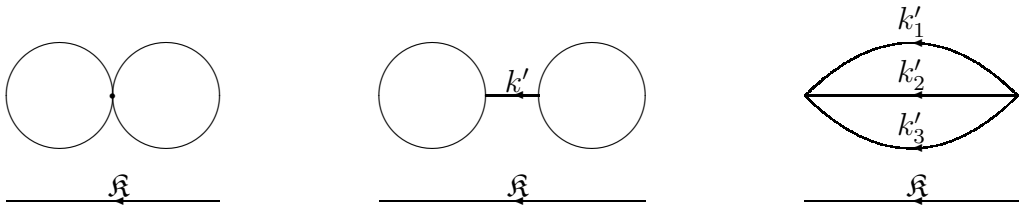

Next we will turn our attention to $E_2$.  Recall from Eq.~(\ref{g2}) that there are two contributions.  The first is equal to $Q_2$ and arises from terms in $H\p\ks$ which contribute to $\Gamma_2^{01}(\kt)$ without ever annihilating the $\bdk$ in $\ks_0$.  In other words, these terms are contained in $\bdk H\p\vac$.   As a result the $\kt$ line is disconnected from the rest of the diagram, which is therefore equivalent to the corresponding diagrams for $H\p\vac$ which were already shown in Fig.~\ref{q2fig}.   These disconnected diagrams are shown in Fig.~\ref{discfig}.

\begin{figure}
\setlength{\unitlength}{.55cm}
\begin{picture}(6,4)(-8,-1)

\put(21.5,-.5){\vl{-1}0 {1.25}}
\put(19,-.5){\vl{-1}0 {1.25}}
\put(20.25,-.2){\makebox(0,0){{$\kt$}}}
\put(17.75,-.2){\makebox(0,0){{$\kt$}}}
\put(19,.5){\circle{2}}

\put(11.5,2){\circle{2}}
\put(12.5,-1){\vector(-1,2){.5}}
\put(12.5,-1){\line(-1,2){1}}
\put(12.5,0){\makebox(0,0){{$k\p$}}}
\put(14,-1){\vl{-1}0 {.75}}
\put(12.5,-1){\vl{-1}0 {1.25}}
\put(13.1,-.7){\makebox(0,0){{$\kt$}}}
\put(11.1,-.7){\makebox(0,0){{$\kt$}}}

\put(7.5,2){\circle{2}}
\put(7.5,1){\vector(-1,-2){.5}}
\put(7.5,1){\line(-1,-2){1}}
\put(7.5,0){\makebox(0,0){{$k\p$}}}
\put(9,-1){\vl{-1}0 {1.25}}
\put(6.5,-1){\vl{-1}0 {0.75}}
\put(7.7,-.7){\makebox(0,0){{$\kt$}}}
\put(5.7,-.7){\makebox(0,0){{$\kt$}}}

\qbezier(-1,0)(0,1)(1,1)
\qbezier(-1,0)(0,0)(1,1)
\put(-.25,.625){\vector(-2,-1)0}
\put(.25,.375){\vector(-2,-1)0}
\put(-.6,.9){\makebox(0,0){{$k\p_1$}}}
\put(.5,.1){\makebox(0,0){{$k\p_2$}}}
\put(2,-1){\vector(-3,1){1.5}}
\put(2,-1){\line(-3,1)3}
\put(.5,-.9){\makebox(0,0){{$\kt$}}}
\put(1,1){\vector(-3,1){1.5}}
\put(1,1){\line(-3,1)3}
\put(-.3,1.8){\makebox(0,0){{$\kt$}}}

\put(-2,0){\vl{-1}0 {1}}
\put(-3,.3){\makebox(0,0){{$\kt$}}}
\put(-6,0){\vl{-1}0 {1}}
\put(-7,.3){\makebox(0,0){{$\kt$}}}
\qbezier(-4,0)(-5,1)(-6,0)
\qbezier(-4,0)(-5,-1)(-6,0)
\put(-5.1,.5){\vector(-1,0)0}
\put(-5.1,-.5){\vector(-1,0)0}
\put(-5,.9){\makebox(0,0){{$k\p_1$}}}
\put(-5,-.9){\makebox(0,0){{$k\p_2$}}}

\end{picture}
\caption{The first two diagrams give the first term in $\mu(\kt)$ as written in Eq.~(\ref{muk}).  The next two are equal and yield the second term.  The last diagram corresponds to the third term.  The other term may be obtained by respectively replacing one $k\p$ in the first two diagrams with a zero mode.}
\label{mufig}
\end{figure}
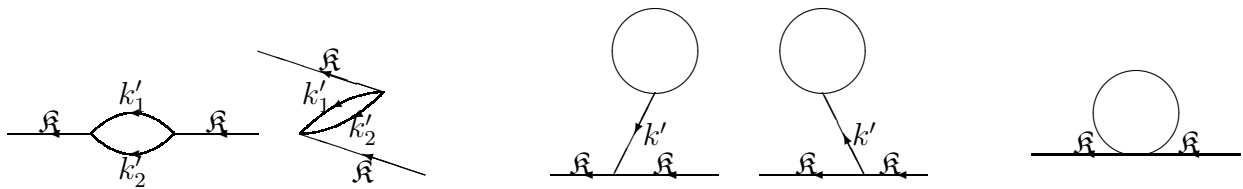

The other contributions to the energy arise from $\mu(\kt)$ in (\ref{muk}).   The first three terms are depicted in Fig.~\ref{mufig}.  Each diagram has a symmetry factor of $1/2$.  Note that in both the third and fourth graphs, the internal line begins at the first (chronologically) vertex and so contributes a factor of $-1/(2\okp{})$.  As a result, the two graphs are equal.  Again graphs for the last two terms are not given.  Intuitively they correspond to the first two graphs with $k\p$ internal lines replaced by zero-mode internal lines.  However again one must choose whether the vertices represent terms in $P\p$ or $H\p$.

\section{Remarks}

We have now found the subleading correction to the normal mode states and their masses.  Are we ready for scattering?

A few more steps are required.  First of all, to calculate matrix elements we will need normalizable states.  These can be made from wave packets of kinks at different momenta.  However, as is, our recursion relation only applies to kinks in the center of mass frame.  The generalization will be straightforward.  Instead of implying that our states are annihilated by $P\p$, we need only impose that they are annihilated by $P\p-p$ for some constant $p$.  This will add a single term to our recursion relation.  As our states are already written in terms of $B^\dag$, $\phi_0$ and $\df$, it will be straightforward to calculate form factors and more general matrix elements with arbitrary polynomials in $\phi(x)$ and $\pi(x)$.  

Next we will need to generalize our results to the interaction picture, as so far we have only considered the Schrodinger picture.  Ideally one would like a suitable Kallen-Lehmann spectral representation and LSZ reduction formula \cite{lsz77,lsz88,dorey93,dorey94,babulsz}.   Also the Wick's theorem of Ref.~\cite{mewick} should be extended to interaction picture fields.

The easiest scattering process to approach would be meson-kink scattering, as this can be done considering the kink state that we have already constructed, adding the perturbative creation operators that create a meson.  We already know their actions on our states, as we have consistently worked in the Fock basis.  If one could prove a factorization theorem in this context, then kink-kink scattering may be treated in some approximation by combining kink-meson scattering with the appropriate form factors.  Here the situation may prove to be much simpler than QCD as the form factors may be calculated perturbatively.

In the quantum theory we expect the phenomenology to be richer than in the classical case.  For example, the normal modes are quantized.  Thus one may expect interesting phenomena, perhaps analogous to \cite{adammultwall}, when integral multiples of the bound normal mode energy pass the threshold $M$ for escape into the continuum.

To efficiently explore such states, it would be useful to complete our construction of a diagrammatic calculus in Sec.~\ref{dessinsez}.   In particular, one should construct rules for $P\p$ vertices in addition to $H\p$ vertices.  In the supersymmetric case, vertices may also represent the supercharges $Q'$.  In the case of rotationally-invariant solitons, vertices may also be introduced for rotations.  

While our perturbative expansion in $P\p$ is much more economical than the exact treatment in the traditional collective coordinate methods of Refs.~\cite{christlee75,gjs}, there is a price to be paid.  As we do not impose that the states are exactly translation-invariant, our solutions are expansions in $\phi_0$ and therefore cease to be reliable if $\phi_0$ is of order $O(1/g)$ corresponding to a kink center of mass position of order $O(1/M)$.  In other words, the kink cannot be coherently treated as its center moves by more than its size.  In the kink rest frame, this is physically reasonable for a semiclassical expansion, it implies that the form factors are dominated by the classical kink solution and quantum corrections are subdominant \cite{jackiw77}.   However in kink scattering it is a limitation, as the kink may never move by $O(1/M)$ in some frame.  This may be an obstruction to constructing an S-matrix, as even scattering with a meson will impart some momentum to the kink which after a time $O(1/(Mg))$ will bring the kink out of this range.

\appendix

\section{Checking $\Gamma_2^{21}$} \label{app}
Recall from (\ref{gcon}) that our state $\ks$ is an eigenstate of the Hamiltonian if it satisfies the Schrodinger equation $\Gamma_i^{mn}=0$.  In the cases $m=0$, at order $i=2$, this condition was imposed by hand to obtain the matrix elements $\gamma_2^{0n}$.  However in Ref.~\cite{me2loop} it was argued that the vanishing of $\Gamma_i^{0n}$, together with translation invariance which was imposed via the recursion relations, is sufficient to make all components vanish.  In this Appendix we will test this claim for the most nontrivial component at order $i=2$, $\Gamma_2^{21}$=0.

We need to calculate all 12 terms that contribute to $\Gamma_2^{21}$.   $(H_2-E_2)\ks_2$ contributes 2 terms, $H_3\ks_1$ contributes 8 terms and $H_4\ks_0$ contribute 2 terms.  We will name these terms $\Gamma_{2,j}^{21}$ and evaluate them one at a time.

The recursion relation yields
\begin{equation}	
		\gamma_{2}^{41}
		=-\frac{\omega_{k_1}\Delta_{k_1 B}\Delta_{-\kt B}}{8}
		-\frac{\gamma_0^{01}(k_1)}{16}\int\frac{dk^{'}}{2\pi}\Delta_{-k^{'}B}\omega_{k^{'}}\Delta_{k^{'}B}.
\end{equation}
This leads to the first contribution
\begin{equation}
	\begin{aligned}
		(H_2-E_1)| \kt \rangle_2\supset
		&\frac{\pi_0^2}{2}\ks_{2}^{41}\\
		=&-6Q_0^{-1}\phi_0^2\int\frac{dk_1}{2\pi} [-\frac{\omega_{k_1}\Delta_{k_1 B}\Delta_{-\kt B}}{8}
		-\frac{\gamma_0^{01}(k_1)}{16}\int\frac{dk\p }{2\pi}\Delta_{-k\p B}\omega_{k\p}\Delta_{k\p B}]\B{1}\vac_0\\ 
	\end{aligned}
\end{equation}
which contributes
\begin{equation}
	\begin{aligned}
		\Gamma_{2,1}^{21}
		=&\frac{3}{4}\omega_{k_1}\Delta_{k_1 B}\Delta_{-\kt B}
		+2\pi\delta(k_1-\kt)\times\frac{3}{8}\int\frac{dk\p }{2\pi}\Delta_{-k\p B}\omega_{k^{'}}\Delta_{k\p B}.
		\\ 
	\end{aligned}
\end{equation}

The other contribution from $(H_2-E_1)\ks_2$ is
\bea
&&(H_2-E_1)\ks_2\supset\left(-\omega_{\kt}+\int\frac{dk}{2\pi}\omega_k B^{\dag}_kB_{k}\right)\ks_{2}^{21}\\
&&=Q_0^{-1}\phi_0^2\int\frac{dk_1}{2\pi}\bigg[\frac{3}{8}\frac{(\wk{1}-\omega_{\kt})^2}{\omega_{\kt}}\Delta_{k_1 B}\Delta_{-\kt B}
-\frac{\sqrt{Q_0}}{8}(\frac{\wk{1}\Delta_{k_1 B}V_{\I-\kt}}{\omega_{\kt}}
+\frac{\omega_{\kt}\Delta_{-\kt B}V_{\I k_1}}{\wk{1}})\nonumber\\
&&+\frac{\sqrt{Q_0}\wk{1}}{8\omega_{\kt}}(\frac{\wk{1}\Delta_{k_1 B}V_{\I-\kt}}{\omega_{\kt}}
+\frac{\omega_{\kt}\Delta_{-\kt B}V_{\I k_1}}{\wk{1}}) \nonumber\\
&&+\frac{1}{8}\int\frac{dk\p_1}{2\pi}\frac{Q_0^{1/2}V_{k_1k\p_1-\kt}}{\omega_{\kt}-\wk{1}-\omega_{k\p_1}}\Delta_{-k\p_1B}
-\frac{1}{8}\int\frac{dk\p_1}{2\pi}\frac{Q_0^{1/2}\wk{1}V_{k_1k\p_1-\kt}}{\omega_{\kt}(\omega_{\kt}-\wk{1}-\omega_{k\p_1})}\Delta_{-k\p_1B}\nonumber\\
&&-\frac{1}{4}\int\frac{dk\p_1}{2\pi}\Delta_{k_1k\p_1}\Delta_{-k\p_1,-\kt}\frac{\wk{1}\omega_{\kt}+\wkp{1}^2}{\wkp{1}}
+\frac{1}{4}\int\frac{dk\p_1}{2\pi}\Delta_{k_1k\p_1}\Delta_{-k\p_1,-\kt}\frac{\wk{1}(\wk{1}\omega_{\kt}+\wkp{1}^2)}{\wkp{1}\omega_{\kt}}\bigg]\
\B{1}\vac_0\nonumber
\eea
yielding
\begin{equation}
	\begin{aligned}
		\Gamma_{2,2}^{21} 
		=&\frac{3}{8}\frac{(\wk{1}-\omega_{\kt})^2}{\omega_{\kt}}\Delta_{k_1 B}\Delta_{-\kt B}
		+\frac{\sqrt{Q_0}}{8}(\frac{\wk{1}^2}{\omega_{\kt}^2}-\frac{\wk{1}}{\omega_{\kt}})\Delta_{k_1 B}V_{\I-\kt}
		+\frac{\sqrt{Q_0}}{8}(1-\frac{\omega_{\kt}}{\wk{1}})\Delta_{\kt B}V_{\I k_1}\\
		&+\frac{1}{8}\int\frac{dk\p_1}{2\pi}\frac{Q_0^{1/2}V_{k_1k\p_1-\kt}}{(\omega_{\kt}-\wk{1}-\omega_{k\p_1})}\Delta_{-k\p_1B}
		-\frac{1}{8}\int\frac{dk\p_1}{2\pi}\frac{Q_0^{1/2}\wk{1}V_{k_1k\p_1-\kt}}{\omega_{\kt}(\omega_{\kt}-\wk{1}-\omega_{k\p_1})}\Delta_{-k\p_1B}\\
		&-\frac{1}{4}\int\frac{dk\p_1}{2\pi}\Delta_{k_1k\p_1}\Delta_{-k\p_1,-\kt}\frac{\wk{1}\omega_{\kt}+\wkp{1}^2}{\wkp{1}}
		+\frac{1}{4}\int\frac{dk\p_1}{2\pi}\Delta_{k_1k\p_1}\Delta_{-k\p_1,-\kt}\frac{\wk{1}(\wk{1}\omega_{\kt}+\wkp{1}^2)}{\omega_{\kt}\wkp{1}}.
	\end{aligned}
\end{equation}

Next we calculate the 8 contributions from $H_3 |\kt\rangle_1$.  The first four arise from
\bea
H_3 |\kt\rangle_1 &\supset&
\frac{3}{6}\int{dx}V^{3}[gf(x)]\phi_0g_B(x)\I(x)\ks_{1}^{11}\\
H_3 |\kt\rangle_1 &\supset&
\frac{1}{6}\int{dx}V^{3}[gf(x)]\int\frac{dk_1}{2\pi}3\phi_0^2 g_B(x)^2g_{k_1}(x)\B{1}\ks_{1}^{00}\nonumber\\
H_3 |\kt\rangle_1 &\supset&
\frac{1}{6}\int{dx}V^{3}[gf(x)]\times\int\frac{dk_1}{2\pi}\frac{1}{2\wk{1}}B_{-k_1}\times 3\phi_0^2g_B^2(x)g_{k_1}(x)\ks_{1}^{02}\nonumber\\
H_3 |\kt\rangle_1 &\supset&
\frac{1}{6}\int{dx}V^{3}[gf(x)]\times\int\frac{dk_1}{2\pi}3\I(x)g_{k_1}(x)\B{1}\ks_{1}^{20}\nonumber
\eea
which respectively contribute
\bea
		\Gamma_{2,3}^{23}
		&=&\frac{Q_0^{1/2}}{4}\frac{\omega_{\kt}+\wk{1}}{\omega_{\kt}}V_{\I B}\Delta_{k_1,-\kt}\\
		\Gamma_{2,4}^{23}
		&=&-\frac{1}{8}\left[\frac{\sqrt{Q_0}\wk{1}^2V_{\I-\kt}}{\omega_{\kt}^2}\Delta_{k_1B}
		+\frac{\wk{1}^2\Delta_{-\kt B}}{\omega_{\kt}}\Delta_{k_1B}\right]		\nonumber\\
		\Gamma_{2,5}^{21} 
		&=&2\pi\delta(k_1-\kt)\frac{1}{8}\int\frac{dk\p}{2\pi}(\sqrt{Q_0}\Delta_{-k\p B}V_{\I   k\p}
		-\omega_{k\p}\Delta_{k\p B}\Delta_{-k\p B})\nonumber\\
		&&+\frac{1}{8}\left(\frac{\sqrt{Q_0}\omega_{\kt}V_{\I k_1}\Delta_{-\kt B}}{\wk{1}}
		-\omega_{\kt}\Delta_{k_1B}\Delta_{-\kt B}\right) 
		-\frac{1}{8}\int\frac{dk\p}{2\pi}\frac{Q_0^{1/2}\omega_{k\p}V_{k_1k\p-\kt}\Delta_{-k\p B}}{\omega_{\kt}(\omega_{\kt}-\wk{1}-\omega_{k\p})}\nonumber\\	
		\Gamma_{2,6}^{21}
		&=&-\frac{Q_0^{1/2}}{8}V_{\I k_1}\Delta_{-\kt B}.	\nonumber
\eea
The other four arise from
\bea
H_3 |\kt\rangle_1&\supset&
\frac{1}{6}\int{dx}V^{3}[gf(x)] \int\frac{dk_1}{2\pi}\frac{3}{2\wk{1}}\I(x)g_{k_1}(x)B_{-k_1}\ks_{1}^{22} \\
H_3 |\kt\rangle_1&\supset&
\frac{1}{6}\int{dx}V^{3}[gf(x)]\times 3\pink{2}\left[\frac{2}{2\wk{2}}\B{1}B_{-k_2}\right]\phi_0 g_B(x)g_{k_1}(x)g_{k_2}(x)\ks_{1}^{11}\nonumber\\
H_3 |\kt\rangle_1& \supset&
\frac{1}{6}\int{dx}V^{3}[gf(x)]\times 3\pink{2}\left[\frac{1}{4\wk{1}\wk{2}}B_{-k_1}B_{-k_2}\right]\phi_0g_B(x)g_{k_1}(x)g_{k_2}(x)\ks_{1}^{13}\nonumber\\
H_3 |\kt\rangle_1 &\supset&
\frac{1}{6}\int{dx}V^{3}[gf(x)]\times\pink{3}\left[\frac{3}{4\wk{2}\wk{3}}\B{1}B_{-k_2}B_{-k_3}\right] \ks_{1}^{22}\nonumber
\eea
and are respectively
\bea
		\Gamma_{2,7}^{21}
		&=&\frac{Q_0^{1/2}}{8}\frac{\wk{1}}{\omega_{\kt}}\Delta_{k_1B}V_{\I -\kt}
		+2\pi\delta(k_1-\kt)\frac{Q_0^{1/2}}{8}\int\frac{dk\p}{2\pi}\Delta_{k\p B}V_{\I -k\p}\\
		\Gamma_{2,8}^{21}
		&=&\frac{1}{4}\int\frac{dk\p}{2\pi}\frac{(\omega_{\kt}+\omega_{k\p})}{\omega_{k\p}\omega_{\kt}}     (\wk{1}^2-\omega_{k\p}^2)\Delta_{k_1k\p}\Delta_{-k\p,-\kt}\nonumber \\
		\Gamma_{2,9}^{21}
		&=&-2\pi\delta(k_1-\kt)\frac{1}{8}\pinkp{2}\frac{(\wkp{1}-\wkp{2})}{\wkp{1}\wkp{2}}
		(\wkp{1}^2-\wkp{2}^2)\Delta_{-k\p_1-k\p_2}\Delta_{k\p_1k\p_2}\nonumber\\
		&&-\frac{1}{4}\int\frac{dk\p}{2\pi}\frac{(\wk{1}-\omega_{k\p})}{\omega_{k\p}\omega_{\kt}}
		(\omega_{k\p}^2-\omega_{\kt}^2)\Delta_{-k\p-\kt}\Delta_{k_1 k\p}\nonumber\\  
		\Gamma_{2,10}^{21}
		&=&\frac{Q_0^{1/2}}{8}\int\frac{dk\p }{2\pi}\frac{V_{k_1k\p-\kt}}{\omega_{\kt}}\Delta_{-k\p B}.  \nonumber
\eea

Finally we arrive at the two contributions from $H_4\ks$.  The first
\beq
H_4 |\kt\rangle_0\supset
\frac{1}{24}\int{dx}V^{4}[gf(x)]\times\pink{2}\frac{2}{2\wk{2}}\B{1}B_{-k_2}\times   6\phi_0^2g_B(x)^2g_{k_1}(x)g_{k_2}(x)B^{\dag}_{\kt}\vac_0
\eeq
contributes
\begin{equation}
	\begin{aligned}
		\Gamma_{2,11}^{21}
		=&\frac{Q_0}{4}\frac{V_{BBk_1-\kt}}{\omega_{\kt}}.
	\end{aligned}
\end{equation}
Using the identity ~\cite{me2loop}
\bea
		V_{BBk_1k_2}&=&\frac{1}{Q_0}\left[-(\wk{1}^2+\wk{2}^2)\Delta_{k_1B}\Delta_{k_2B}\right.\\
		&&+\int\frac{dk^{'}}{2\pi}(-\sqrt{Q_0}V_{k_1k_2k^{'}}\Delta_{-k^{'}B}
		\left.+(\wk{1}^2+\wk{2}^2-2\omega_{k\p}^2)\Delta_{k_2k\p}\Delta_{-k\p k_1})\right]\nonumber
\eea
this can be written
\bea
		\Gamma_{2,11}^{21}
		&=&-\frac{1}{4\omega_{\kt}}(\wk{1}^2+\omega_{\kt}^2)\Delta_{k_1B}\Delta_{-\kt B}
		-\frac{\sqrt{Q_0}}{4\omega_{\kt}}\int\frac{dk\p}{2\pi}V_{k_1-\kt k\p}\Delta_{-k\p B}\\
		&&		+\int\frac{dk\p}{2\pi}\frac{(\wk{1}^2+\omega_{\kt}^2-2\omega_{k\p}^2)}{4\omega_{\kt}}\Delta_{-\kt k\p}\Delta_{-k\p k_1}.\nonumber
\eea
The last contribution arises from
\beq
H_4 |\kt\rangle_0\supset
\frac{1}{24}\int{dx}V^{4}[gf(x)]6\I(x) \phi_0^2g_B(x)^2\times  B^{\dag}_{\kt}\vac_0
\eeq
and is equal to
\begin{equation}
	\begin{aligned}
		\Gamma_{2,12}^{21}
		=&2\pi\delta(k_1-\kt)\frac{Q_0}{4}V_{\I BB}.
	\end{aligned}
\end{equation}
The identity~\cite{me2loop}
\begin{equation}
	\begin{aligned}
		V_{\I BB}=\frac{1}{Q_0}\left(\pinkp{2}\frac{\wkp{1}^2-\wkp{2}^2}{\wkp{1}}\Delta_{k\p_1k\p_2}\Delta_{-k\p_1-k\p_2}
		+\int\frac{dk\p}{2\pi}\omega_{k\p}\Delta_{Bk\p}\Delta_{-k\p B}
		-\sqrt{Q_0}\int\frac{dk\p}{2\pi}V_{\I k\p}\Delta_{-k\p B}\right)	\\    
	\end{aligned}
\end{equation}
again allows terms involving the integrals of four $g(x)$ to be eliminated, leaving
\begin{equation}
	\begin{aligned}
		\Gamma_{2,12}^{21}
		=&\frac{2\pi\delta(k_1-\kt)}{4}\bigg(\pinkp{2}\frac{\wkp{1}^2-\wkp{2}^2}{\wkp{1}}\Delta_{k\p_1k\p_2}
		\Delta_{-k\p_1-k\p_2}
		+\int\frac{dk\p}{2\pi}\omega_{k\p}\Delta_{Bk\p}\Delta_{-k\p B}\\
		&-\sqrt{Q_0}\int\frac{dk\p}{2\pi}V_{\I k\p}\Delta_{-k\p B}\bigg).\\    
	\end{aligned}
\end{equation}
Without these identities we would not be able to show that $\Gamma=0$.

Finally, summing all of the above contributions, we obtain
\begin{equation}
	\begin{aligned}
		&\Gamma_2^{21}=\sum_{i=1}^{12}\Gamma_{2,i}^{21}=A(k_1)+2\pi\delta(k_1-\kt)B(k_1)+\int \frac{dk^{'}}{2\pi}C(k_1)     .                                
	\end{aligned}
\end{equation}
The first term is
\begin{equation}
	\begin{aligned}
		A(k_1)=&\frac{3}{4}\wk{1}\Delta_{k_1 B}\Delta_{-\kt B}
		+\frac{3}{8}\frac{(\wk{1}-\omega_{\kt})^2}{\omega_{\kt}}\Delta_{k_1 B}\Delta_{-\kt B}
		+\frac{\sqrt{Q_0}}{8}(\frac{\wk{1}^2}{\omega_{\kt}^2}-\frac{\wk{1}}{\omega_{\kt}})\Delta_{k_1 B}V_{\I-\kt}\\
		&+\frac{\sqrt{Q_0}}{8}(1-\frac{\omega_{\kt}}{\wk{1}})\Delta_{\kt B}V_{\I k_1}
		+\frac{Q_0^{1/2}}{4}\frac{\omega_{\kt}+\wk{1}}{\omega_{\kt}}V_{\I B}\Delta_{k_{1},-\kt}
		-\frac{1}{8}[\frac{\sqrt{Q_0}\wk{1}^2V_{\I-\kt}}{\omega_{\kt}^2}\Delta_{k_1B}\\
		&+\frac{\wk{1}^2\Delta_{-\kt B}}{\omega_{\kt}}\Delta_{k_1B}]
		-\frac{1}{4\omega_{\kt}}(\wk{1}^2+\omega_{\kt}^2)\Delta_{k_1B}\Delta_{-\kt B}
		+\frac{1}{8}(\frac{\sqrt{Q_0}\omega_{\kt}V_{\I k_1}\Delta_{-\kt B}}{\wk{1}}\\
		&-\omega_{\kt}\Delta_{k_1B}\Delta_{-\kt B})
		-\frac{Q_0^{1/2}}{8}V_{\I k_1}\Delta_{-\kt B}
		+\frac{Q_0^{1/2}\wk{1}}{8\omega_{\kt}}V_{\I -\kt}\Delta_{k_1 B}\\ 
		=&0
	\end{aligned}
\end{equation}
where use the fact~\cite{me2loop} that $V_{\I B}=0$.  The other terms are
\begin{equation}
	\begin{aligned}
		B(k_1)=&\frac{3}{8}\int\frac{dk\p}{2\pi}\Delta_{-k\p B}\omega_{k\p}\Delta_{k\p B}
		+\frac{1}{8}\int\frac{dk\p}{2\pi}(\sqrt{Q_0}\Delta_{-k\p B}V_{\I   k\p}
		-\omega_{k\p}\Delta_{k\p B}\Delta_{-k\p B}+\sqrt{Q_0}\Delta_{k\p B}V_{\I   -k\p})\\
		&-\frac{1}{8}\pinkp{2}\frac{(\wkp{1}-\wkp{2})}{\wkp{1}\wkp{2}}(\wkp{1}^2-\wkp{2}^2)\Delta_{-k\p_1-k\p_2}
		\Delta_{k\p_1k\p_2}\\
		&+\frac{1}{4}\left(\pinkp{2}\frac{\wkp{1}^2-\wkp{2}^2}{\wkp{1}}\Delta_{k\p_1k\p_2}\Delta_{-k\p_1-k\p_2}
		+\int\frac{dk\p}{2\pi}\omega_{k\p}\Delta_{Bk\p}\Delta_{-k\p B}
		-\sqrt{Q_0}\int\frac{dk\p}{2\pi}V_{\I k\p}\Delta_{-k\p B}\right)\\
		=0
	\end{aligned}
\end{equation}
and
\begin{equation}
	\begin{aligned}
		C(k_1)=&\frac{1}{8}\frac{Q_0^{1/2}V_{k_1k\p_{1}-\kt}}{(\omega_{\kt}-\wk{1}-\wkp{1})}\Delta_{-k\p B}
		-\frac{1}{8}\frac{Q_0^{1/2}\wk{1}V_{k_1k\p-\kt}}{\omega_{\kt}(\omega_{\kt}-\wk{1}-\omega_{k\p})}\Delta_{-k\p B}
		-\frac{1}{4}\Delta_{k_1k\p}\Delta_{-k\p,-\kt}\frac{\wk{1}\omega_{\kt}+\omega_{k\p}^2}{\omega_{k\p}}\\
		&+\frac{1}{4}\Delta_{k_1k\p}\Delta_{-k\p,-\kt}\frac{\wk{1}(\wk{1}\omega_{\kt}+\omega_{k\p}^2)}{\omega_{\kt}\omega_{k\p}} 
		-\frac{1}{8}\frac{Q_0^{1/2}\omega_{k\p}V_{k_1k\p-\kt}\Delta_{-k\p B}}{\omega_{\kt}(\omega_{\kt}-\wk{1}-\omega_{k\p})}\\	 
		&-\frac{1}{4}\frac{(\omega_{\kt}+\omega_{k\p})}{\omega_{k\p}\omega_{\kt}}(\wk{1}^2-\omega_{k\p}^2)\Delta_{k_1k\p}\Delta_{-k\p,-\kt} 
		-\frac{1}{4}\frac{(\wk{1}-\omega_{k\p})}{\omega_{k\p}\omega_{\kt}}(\omega_{k\p}^2-\omega_{\kt}^2)
		\Delta_{-k\p-\kt}\Delta_{k_1 k\p}\\
		&+\frac{Q_0^{1/2}}{8}\frac{V_{k_1k\p-\kt}}{\omega_{\kt}}\Delta_{-k\p B}
		-\frac{\sqrt{Q_0}}{4\omega_{\kt}}V_{k_1-\kt k\p}\Delta_{-k\p B}
		+\frac{(\wk{1}^2+\omega_{\kt}^2-2\omega_{k\p}^2)}{4\omega_{\kt}}\Delta_{-\kt -k\p}\Delta_{k\p k_1}\\
		=0.\\  
	\end{aligned}
\end{equation}
As all three contributions vanish, we have shown that
\beq
\Gamma_2^{21}=0
\eeq
as it must be if $\ks$ is indeed a Hamiltonian eigenstate to second order.
\section* {Acknowledgement}

\noindent
JE is supported by the CAS Key Research Program of Frontier Sciences grant QYZDY-SSW-SLH006 and the NSFC MianShang grants 11875296 and 11675223.   JE also thanks the Recruitment Program of High-end Foreign Experts for support.

\end{document}